\documentclass[12pt]{article}
\usepackage[english]{babel}
\usepackage[utf8]{inputenc}
\usepackage[T1]{fontenc}

\def\tmH{\tilde{\mH}}
\def\mF{\mathcal{F}}
\usepackage{amsmath}

\usepackage{amsfonts}
\def\mU{\mathcal{U}}
\def\bH{\mathbf{H}}
\usepackage{url}
\usepackage[dvips]{graphicx}
\usepackage{calc}
\usepackage{subfigure}
\usepackage{multirow}

\def\mL{\mathcal{L}}
\def\bD{\mathbf{D}}
\def\mH{\mathcal{H}}
\def\mD{\mathcal{D}}
\def\bx{\mathbf{x}}
\newcommand{\pb}[1]{\left\{#1\right\}}

\def\tX{\tilde{X}}

\def\mG{\mathcal{G}}
\def\by{\mathbf{y}}

\def\bT{\mathbf{T}}

\def\mM{\mathcal{M}}

\begin{document}
	\begin{center}
		{\Large{ \bf Hamiltonian for  Weyl Transverse Gravity}}
		
		\vspace{1em}  J. Kluso\v{n} 
		\footnote{Email addresses:
			J. Kluso\v{n}:\ klu@physics.muni.cz} \\
		\vspace{1em}   \textit{Department of Theoretical Physics and
			Astrophysics, Faculty
			of Science,\\
			Masaryk University, Kotl\'a\v{r}sk\'a 2, 611 37, Brno, Czech Republic}\\

	\end{center}
	
	\abstract{In this short note we determine Hamiltonian for Weyl transverse gravity. 
	We find primary, secondary and tertiary constraints and calculate Poisson brackets 
between them. We also show that gauge fixing in Weyl transverse gravity leads to the Hamiltonian 
for unimodular gravity. }
	
\newpage

\section{Introduction and Summary}\label{first}
Diffeomorphism invariance is fundamental postulate of theoretical physics that says that the basic laws of nature should be formulated using action functional which is manifestly invariant under diffeomorphism transformations. On the other hand we can still keep in mind that nature at fundamental level does not have to respect this presumption. Famous example of  such proposal is Ho\v{r}ava-Lifshitz gravity \cite{Horava:2009uw} where the action has manifestly non-covariant form and it is invariant under restricted diffeomorphism
transformations.  Another example of such a theory  is unimodular gravity that is invariant under transverse diffeomorphism. 
Unimodular gravity was firstly introduced by A. Einstein in his paper 
\cite{Einstein:1916vd} that was published in 1916. In this work the unimodular constraint
$\sqrt{-g}=1$ was used as gauge fixing condition 
of general diffeomorphism in order to simplify calculations. Then it was shown in 
\cite{Buchmuller:1988wx,Buchmuller:1988yn} that imposing this condition before the variation of Einstein-Hilbert action leads to the traceless equations of motion. An attractive property of 
this theory is that the cosmological constant appears as an integration constant which varies from solution to solution. Similar property possesses also generalization of unimodular gravity which is Weyl transverse gravity (WTG) \cite{Oda:2016psn}. This theory is invariant under diffeomorphism and Weyl transformations
\footnote{For recent works related to Weyl transverse gravity, see 
\cite{Alonso-Serrano:2023gum,Kluson:2023idq,Alonso-Serrano:2022pif,Oda:2022txb,
Alonso-Serrano:2022rzj,Gielen:2018pvk}, see also 
\cite{Kugo:2021bej,Kugo:2022iob,Kugo:2022dui}.}. An important property of Weyl transverse gravity is that it is an alternative to General relativity and arises from the observation that consistent
theory of self-interacting gravitons does not uniquely lead to General relativity. Instead 
such a theory could be invariant under transverse diffeomorphism and Weyl transformations
\cite{Alvarez:2006uu,Barcelo:2014mua}.

The goal of this paper is to find canonical formulation of Weyl transverse gravity. This is non-trivial task due to the presence of the kinetic term for determinant of metric which 
appears in the action for Weyl transverse gravity. Canonical analysis of theories with kinetic term for determinant of metric was performed recently in \cite{Kluson:2023gnh,Kluson:2023bkg}.
We formulated these theories with the help of the variables that define $(n-1)+1$-formalism of gravity in $n$ dimensions, for review see \cite{Gourgoulhon:2007ue}. We showed recently 
in \cite{Kluson:2023gnh} that this theory does not have well defined canonical structure due to the fact that Hamiltonian constraint is the second class constraint. We mean that reason for the failure of canonical formalism was in the fact that we worked in specific coordinates where
the volume form was constant and equal to one. In fact, in this paper we will keep $\omega$ as a background volume form arbitrary. Then  WTG  does not have kinetic term for $g\equiv \det g_{\mu\nu}$ but instead it depends on the combination $\frac{\sqrt{-g}}{\omega}$ which behaves as scalar under diffeomorphism transformations. Then it is natural to introduce auxiliary scalar field that is equal to $\sqrt{-g}/\omega$ on shell. Introducing this scalar field we will be able to rewrite WTG action that is suitable for the canonical formalism. Using this action we find corresponding Hamiltonian and we identify primary constraints of the theory. Then we will study stability of these constraints and we show that new set of spatial diffeomorphism constraints is generated. As the final step we should study stability of these constraints and
we find that new, tertiary constraint is generated. More precisely, this constraint has the form of derivative of Hamiltonian function and we argue that the proper solution of this 
condition is set of $(n-1)\times \infty-1$
\footnote{We used notation introduced in \cite{Kuchar:1991xd}.}
 Hamiltonian constraints together with zero mode part of Hamiltonian which is not restricted. 
This result is the same as in the case of the unimodular gravity as was shown in 
\cite{Kuchar:1991xd} and  \cite{Henneaux:1989zc}. In fact, the zero mode part of Hamiltonian 
does not change under time evolution of system and effectively corresponds to cosmological constant which is again the same result as in unimodular gravity and which is natural consequence of the fact that fixing the Weyl symmetry of WTG leads to unimodular gravity as we show in the end of this paper. 

Let us outline our results. We found canonical formulation of Weyl transverse gravity. We explicitly identified primary, secondary and tertiary constraints and we calculated Poisson brackets between them. Finally we showed that gauge fixing of Weyl symmetry naturally led to Hamiltonian for unimodular gravity. 

This paper is organized as follows. In the next section (\ref{second}) we introduce WTG action and determine corresponding Hamiltonian. Then in section (\ref{third}) we study time evolutions of constraints and finalize analysis of canonical structure of theory.

\section{Weyl Transverse Gravity}\label{second}
In this section we find canonical form of Weyl transverse gravity.  WTG action  in $n-$dimensions was introduced in \cite{Oda:2016psn} and it takes the form 
\begin{eqnarray}\label{actWeyl}
S=\frac{1}{\kappa}\int d^nx
\omega
\left(\frac{\sqrt{-g}}{\omega}\right)^{\frac{2}{n}}
\left[R+\frac{(n-1)(n-2)}{n^2}
g^{\mu\nu}\partial_\mu \ln (\frac{\sqrt{-g}}{\omega})
\partial_\nu\ln (\frac{\sqrt{-g}}{\omega})\right]
 \ , 
\nonumber \\
\end{eqnarray}
where $g=\det g_{\mu\nu}$ and and where $\omega(x)$ is volume form. The presence of volume form means that the volume element $d^nx \omega(x)$ is invariant under diffeomorphism transformations and also that  $\frac{\sqrt{-g}}{\omega}$ behaves as scalar under diffeomorphism transformations. 

In order to find Hamiltonian for WTG we should carefully examine the kinetic term for the metric $g$. In our previous papers \cite{Kluson:2023gnh,Kluson:2023bkg}
we evaluated this term explicitly using $(n-1)-1$ decomposition of the metric. However thanks to the presence of non-dynamical volume form $\omega(x)$ that makes the action manifestly invariant under diffeomorphism transformations it is much more convenient to introduce new scalar field $\phi$ and rewrite the action
 (\ref{actWeyl}) into  an equivalent form 
\begin{eqnarray}\label{actWeyl2}
&&S=\frac{1}{\kappa}\int d^nx
\omega\phi^{\frac{2}{n}}\left[R+\frac{(n-1)(n-2)}{n^2\phi^2}
g^{\mu\nu}\partial_\mu\phi\partial_\nu\phi+
\lambda\left(\phi-\frac{\sqrt{-g}}{\omega}\right)
\right] \ , 
\nonumber \\
\end{eqnarray}
where $\lambda$ is Lagrange multiplier so that the variation of the action with respect to $\lambda$
gives $\phi=\frac{\sqrt{-g}}{\omega}$ and hence it is clear that the action (\ref{actWeyl2}) 
is equivalent to the action (\ref{actWeyl}). In what follows we will work with the action 
(\ref{actWeyl2}) in order to find canonical formulation of Weyl transverse gravity. 

Standard treatment of canonical formulation of gravity is based on $(n-1)+1$ splitting 
of space-time  \footnote{For recent review, see
	\cite{Gourgoulhon:2007ue}.}. In more details, we consider $n$ dimensional manifold
$\mathcal{M}$ with the coordinates $x^\mu \ , \mu=0,\dots,n-1$ and
where $x^\mu=(t,\bx) \ , \bx=(x^1,x^2,\dots, x^{n-1})$. We presume that this
space-time is endowed with the metric $g_{\mu\nu}(x^\rho)$
with signature $(-,+,\dots,+)$. Suppose that $ \mathcal{M}$ can be
foliated by a family of space-like surfaces $\Sigma_t$ defined by
$t=x^0=\mathrm{const}$. Let $h_{ij}, i,j=1,2,\dots,n-1$ denotes the metric on $\Sigma_t$
with inverse $h^{ij}$ so that $h_{ij}h^{jk}= \delta_i^k$. We further
introduce the operator $\nabla_i$ that is covariant derivative
defined with the metric $h_{ij}$.
We also define  the lapse
function $N=1/\sqrt{-g^{00}}$ and the shift function
$N^i=-g^{0i}/g^{00}$. In terms of these variables we
write the components of the metric $g_{\mu\nu}$ as
\begin{eqnarray}
&&g_{00}=-N^2+N_i h^{ij}N_j \ , \quad g_{0i}=N_i \ , \quad
g_{ij}=h_{ij} \ ,
\nonumber \\
&&g^{00}=-\frac{1}{N^2} \ , \quad g^{0i}=\frac{N^i}{N^2} \
, \quad g^{ij}=h^{ij}-\frac{N^i N^j}{N^2} \ 
\nonumber \\
\end{eqnarray}
and hence $g=-N^2\det h_{ij}$. We  also have following decomposition of $R$ in the form
\begin{eqnarray}
&&R=K^{ij}K_{ij}-K^2+r+2\tilde{\nabla}_\mu[\tilde{n}^\mu K]
-\frac{2}{N}\nabla_i\nabla^iN \ , 
\nonumber \\
&&K_{ij}=\frac{1}{2N}(\partial_t h_{ij}-\nabla_i N_j-\nabla_j N_i) \ , 
\quad n^0=\sqrt{-g^{00}} \ , \quad 
n^i=-\frac{g^{0i}}{\sqrt{-g^{00}}} \ , \nonumber \\
\end{eqnarray}
and where $r$ is scalar curvature defined with $h_{ij}$ and where $\tilde{\nabla}_\mu$ is covariant derivative compatible with $g_{\mu\nu}$ while 
$\nabla_i$ is covariant derivative compatible with the metric $h_{ij}$. Note that we can also write
\begin{eqnarray}
&&\tilde{\nabla}_\mu [n^\mu K]=\frac{1}{\sqrt{-g}}\partial_\mu[\sqrt{-g}n^\mu K]
\ , \nonumber \\
&&g^{\mu\nu}\partial_\mu\phi\partial_\nu\phi=-D_n\phi^2+h^{ij}\partial_i\phi
\partial_j\phi \ , \quad  D_n\phi=\frac{1}{N}(\partial_0\phi-N^i\partial_i\phi) \ . 
\nonumber \\
\end{eqnarray}
Using this notation the action (\ref{actWeyl2}) has the form 
\begin{eqnarray}\label{Sn}
&&S=\frac{1}{\kappa}\int d^nx \omega
\phi^{\frac{2}{n}}\left[
K_{ij}\mG^{ijkl}K_{kl}+ r-\frac{2(2-n)}{n}\frac{1}{\phi}D_n\phi K
-\frac{2}{\phi}\partial_i[\sqrt{h}h^{ij}\partial_j N]\right.
-\nonumber \\
&&\left.-\frac{(n-1)(n-2)}{n^2\phi^2}
D_n\phi^2+\frac{(n-1)(n-2)}{\phi^2}h^{ij}\partial_i\phi
\partial_j\phi+ \lambda\left(\phi-\frac{\sqrt{-g}}{\omega}\right)\right] \ , 
\nonumber \\
\end{eqnarray}
where we ignored boundary terms and 
where we replaced $\frac{\omega}{\sqrt{-g}}$ with $\frac{1}{\phi}$. 
Further,  Wheeler–DeWitt metric $\mG^{ijkl}$ is defined as
\begin{equation}
\mG^{ijkl}=\frac{1}{2}(h^{ik}h^{jl}+h^{il}h^{jk})-
h^{ij}h^{kl} \ 
\end{equation}
with inverse 
\begin{equation}\label{mGinv}
\mG_{ijkl}=\frac{1}{2}(h_{ik}h_{jl}+h_{il}h_{jk})-
\frac{1}{n-2}h_{ik}h_{jl} \ . 
\end{equation}
In order to find
canonical formulation it is convenient to rewrite the action (\ref{Sn}) into the form 
with quadratic kinetic terms 
\begin{eqnarray}\label{Squadr}
&&S=\frac{1}{\kappa}\int d^nx 
\omega \phi^{\frac{2}{n}}\left[K_{ij}\mM^{ijkl}K_{kl}-
\frac{(n-1)(n-2)}{n^2}\tX^2+r-\frac{2}{\phi}\partial_i[\sqrt{h}h^{ij}
\partial_jN]+\right.\nonumber \\
&&\left.+\frac{(n-1)(n-2)}{\phi^2}h^{ij}\partial_i\phi\partial_j\phi+
\lambda\left(\phi-\frac{\sqrt{-g}}{\omega}\right)\right] \ , 
\nonumber \\
\end{eqnarray}
where $\mM^{ijkl}$ and $\tX$ are defined as 
\begin{eqnarray}\label{defmM}
&&\mM^{ijkl}=\mG^{ijkl}+\frac{n-2}{n-1}h^{ij}h^{kl}=
\frac{1}{2}(h^{ik}h^{jl}+h^{il}h^{jk})-\frac{1}{n-1}h^{ij}h^{kl} 
\ , 
\nonumber \\
&&\tX=\frac{D_n\phi}{\phi}-\frac{n}{n-1}K \ . \nonumber \\
\end{eqnarray}
From the action (\ref{Squadr}) we obtain conjugate momenta
\begin{eqnarray}\label{momenta}
&&\pi^{ij}=\frac{\partial \mL}{\partial(\partial_0 h_{ij})}=
\frac{\omega}{\kappa}\phi^{\frac{2}{n}}\left[\frac{1}{N}\mM^{ijkl}K_{kl}
+\frac{(n-2)}{nN}h^{ij}\tX\right] \ , \nonumber \\
&&p_\phi=\frac{\partial \mL}{\partial(\partial_0\phi)}=
-2\frac{\omega}{\kappa}\phi^{\frac{2}{n}} \frac{(n-1)(n-2)}{n^2}\frac{1}{\phi N}\tX \ . 
\nonumber \\
\end{eqnarray}
Inserting the second relation (\ref{momenta}) into the first one we obtain 
\begin{equation}\label{relPimM}
\Pi^{ij}=\frac{\omega}{\kappa}\phi^{\frac{2}{n}} 
\mM^{ijkl}K_{kl} \ , 
\end{equation}
where we defined $\Pi^{ij}$ as 
\begin{equation}
\Pi^{ij}=\pi^{ij}+
	\frac{n}{2(n-1)}p_\phi \phi h^{ij} \ . 
\end{equation}
It is important to stress that it is not possible
to express $K_{ij}$ as
function of $\Pi^{ij}$
 from (\ref{relPimM})    due to the fact that $\mM^{ijkl}$ obeys following relations
\begin{equation}\label{mMrel}
h_{ij}\mM^{ijkl}=0 \ , \quad 
\mM^{ijkl}h_{km}h_{ln}\mM^{mnrs}=\mM^{ijrs}
\end{equation}
as follows from the explicit form of $\mM^{ijkl}$ given in (\ref{defmM}). In fact, 
using the first relation in (\ref{relPimM}) we obtain following primary constraint
\begin{equation}
h_{ij}\Pi^{ij}=\pi+\frac{n}{2}p_\phi \phi\equiv \mD\approx 0 \ . 
\end{equation}
Further, using the second relation given in (\ref{mMrel}) we can write
$K_{ij}\mM^{ijkl}K_{kl}=K_{ij}\mM^{ijkl}h_{km}h_{ln}\mM^{mnrs}K_{rs}$ and then 
using (\ref{relPimM}) we can express Hamiltonian as function of canonical variables. 
In more details, the Hamiltonian is defined as 
\begin{eqnarray}
&&H=\int d^{n-1}\bx(\partial_0 h_{ij}\pi^{ij}+p_\phi \partial_0\phi-\mL)=
\nonumber \\
&&=\int d^{n-1}\bx(2NK_{ij}\pi^{ij}+Np_\phi D_n\phi+2\nabla_iN_j\pi^{ij}+N^ip_\phi
\partial_i\phi-\mL)=\nonumber \\
&&=\int d^{n-1}\bx\left[(\frac{\kappa}{\omega})\phi^{-\frac{2}{n}}
N^2\Pi^{ij}h_{ik}h_{jl}\Pi^{kl}
-\frac{\kappa}{4\omega}\frac{n^2}{(n-1)(n-2)}\phi^{-\frac{2}{n}}\phi^2N^2p_\phi^2
-\right.\nonumber \\
&&-\frac{\omega}{\kappa}\phi^{\frac{2}{n}}r+
\frac{2}{\kappa}N\partial_i[\sqrt{h}h^{ij}\partial_j\phi^{\frac{2}{n}-1}]
-\frac{\omega}{\kappa}\phi^{\frac{2}{n}}
\frac{(n-1)(n-2)}{n^2\phi^2}h^{ij}\partial_i\phi\partial_j\phi+\nonumber \\
&&\left.+N^i\mH_i+
\frac{1}{\kappa}\lambda\omega\phi^{\frac{2}{n}}\left(\phi-\frac{\sqrt{-g}}{\omega}\right)\right]  \ , 
\nonumber \\
\end{eqnarray}
where
\begin{equation}
\mH_i=-2h_{ik}\nabla_j \pi^{kj}+p_\phi \partial_i\phi \ . 
\end{equation}
It turns out that it is convenient to  to express $N$ as $N=\frac{\phi \omega}{\sqrt{h}}$ so that the Hamiltonian is equal to 
\footnote{Note that this is certainly correct procedure as we demonstrate on following example. Let us have an action functional for dynamical variable 
$A$ in the form $\int A^\alpha [g^{\mu\nu}\partial_\mu A\partial_\nu A+V(A)]$. This action is on-shell equivalent to the following action $\int( A^{\alpha-\beta}B^\beta[g^{\mu\nu}\partial_\mu B\partial_\nu B+V(B)]+\lambda(A-B))$ where $B$ is new variable, $\lambda$ is Lagrange multiplier and where $\beta$ is arbitrary number. }
\begin{eqnarray}\label{HBfinal}
	H_B
	\int d^{n-1}\bx (\omega \tmH_0+N^i\mH_i+\lambda \omega \phi^{\frac{2}{n}}(\phi-\frac{N\sqrt{h}}{\omega})+
	v_N\pi_N+v^i\pi_I+v^\lambda p_\lambda+\Sigma \mD) \ , 
	\nonumber \\
\end{eqnarray} 
where
\begin{eqnarray}
&&	\tmH_0=	\kappa \phi^{-\frac{2}{n}}\frac{\phi^2}{h}\Pi^{ij}h_{ik}h_{jl}\Pi^{kl}
	-\frac{\kappa}{4}\frac{n^2}{(n-1)(n-2)}
	\phi^{-\frac{2}{n}}\frac{\phi^4}{h}p_\phi^2-\nonumber \\
&&	-\frac{1}{\kappa }\phi^{\frac{2}{n}}r+\frac{2\phi}{\kappa\sqrt{h}}
	\partial_i[\sqrt{h}h^{ij}\partial_j\phi^{(\frac{2}{n}-1)}]
	-\frac{1}{\kappa}\phi^{\frac{2}{n}}\frac{(n-1)(n-2)}{n^2\phi^2}
	h^{ij}\partial_i\phi\partial_j\phi \ , 
	\nonumber \\ 
\end{eqnarray}
where we also performed integration by parts and 
where $v_N,v^i,v^\lambda$ and $\Sigma$ are Lagrange multipliers corresponding to the primary constraints 
\begin{equation}\label{primcon}
\pi_N\approx 0 \ , \quad 	\pi_i\approx 0 \ ,  \quad  p_\lambda \approx 0 \ , \quad \mD\approx 0 \ . 
\end{equation}
In what follows we will work with the form of the Hamiltonian given above. As the first step we analyse 
time evolution of primary constraints.

\section{Preservation of primary constraints}\label{third}
Theory with constraints is well defined when the primary constraints
are preserved during the time evolution of system. In case of   Weyl transverse gravity these constraints are given in (\ref{primcon}) and now we will study their preservation. 
The time evolution of any phase-space function is given by its Poisson bracket with the bare Hamiltonian $H_B$. Note that we have following non-zero canonical Poisson
brackets
\begin{eqnarray}\label{canpb}
&&\pb{h_{ij}(\bx),\pi^{kl}(\by)}=\frac{1}{2}(\delta_i^k\delta_j^l+\delta_i^l
\delta_j^k)\delta(\bx-\by) \ , \pb{N^i(\bx),\pi_j(\by)}=\delta^i_j\delta(\bx-\by) \ , 
\nonumber \\
&&\pb{N(\bx),\pi_N(\by)}=\delta(\bx-\by)  \ , \quad 
\pb{\phi(\bx),p_\phi(\by)}=\delta(\bx-\by) \ , \nonumber \\
&&
\pb{\lambda(\bx),p_\lambda(\by)}=\delta(\bx-\by) \  \quad 
\nonumber \\
\end{eqnarray}
so that time evolution of the constraint  $\pi_i\approx 0$ has the form
\begin{equation}
	\partial_0\pi_i=\pb{\pi_i,H_B}=-\mH_i
\end{equation}
and we see that the constraint $\pi_i\approx 0$ will be preserved when we impose secondary constraints
\begin{equation}
\mH_i\approx 0 \ . 
\end{equation} 
In case of constraint $p_\lambda \approx 0$ we obtain
\begin{equation}
	\partial_0 p_\lambda=\pb{p_\lambda,H_B}=-\frac{1}{\kappa}\omega \phi^{\frac{2}{n}}
	\left(\phi-\frac{N\sqrt{h}}{\omega}\right) \equiv -\omega \phi^{\frac{2}{n}}
	\mU\approx 0 \ ,
\end{equation}
where again we should impose secondary constraint $\mU\approx 0$. Finally 
 time evolution of the constraint $
\pi_N\approx 0$  has the form 
\begin{equation}
	\partial_0\pi_N=\pb{\pi_N,H_B}=
	\frac{1}{\kappa}\omega\lambda \phi^{\frac{2}{n}}\sqrt{h} 
\end{equation}
and hence requirement of the preservation of the constraint $\pi_N\approx 0$  leads to another secondary constraint
\begin{equation}
\lambda \approx 0 \ . 
\end{equation}
  In fact, we see that the constraints
$p_\lambda$ and $\lambda$ decouple from the rest of the system. Finally the requirement of the preservation of constraint $\mU$ implies
\begin{equation}
	\partial_t\mU=\pb{\mU,H_B}=\int d^{n-1}\by\omega\pb{\mU,\tmH_0(\by)}+\frac{\sqrt{h}}{\omega}v_N=0
\end{equation}
that can be solved for $v_N$ which is a consequence of the fact that  constraints $\pi_N\approx 0$ and $\mU\approx 0$ are second class constraints. Since second class constraints vanish strongly they  can be solved for $N$ and $\pi_N$ as $\pi_N=0$ and $N=\frac{\phi \omega}{\sqrt{h}}$. In other words $\pi_N$ and $N$ decouple from the rest of the system as well.


Finally we study requirement of the preservation of the constraint $\mD\approx 0$. It is convenient to introduce its smeared form 
\begin{equation}
	\bD=\int d^{n-1}\bx \Omega(\bx)\mD(\bx) \ . 
\end{equation}
Then using canonical Poisson brackets (\ref{canpb}) 
it is easy to determine following Poisson brackets 
\begin{eqnarray}
&&	\pb{\bD(\Omega),h_{ij}(\by)}=-\Omega h_{ij}(\by) \ , \quad 
	\pb{\bD(\Omega),\pi^{ij}(\by)}=\Omega\pi^{ij} (\by) \ , \nonumber \\
&&	\pb{\bD(\Omega),\phi(\by)}=-\frac{n}{2}\Omega \phi(\by) \ , \quad 
	\pb{\bD(\Omega),p_\phi(\by)}=\frac{n}{2}\Omega p_\phi(\by) \ ,\nonumber \\
&&	\pb{\bD(\Omega),\Pi^{ij}(\by)}=\Omega \Pi^{ij}(\by) \ , \quad 
	\pb{\bD(\Omega),\sqrt{h}(\by)}=-\frac{(n-1)}{2}\Omega\sqrt{h}(\by) \  . \nonumber \\
\end{eqnarray}
Then we can calculate   
\begin{eqnarray}
&&	\pb{\bD(\Omega),-\int d^{n-1}\by \frac{\sqrt{h}}{\kappa\phi}\phi^{\frac{2}{n}}r}=
	-\frac{1}{2}\int d^{n-1}\by \Omega \frac{\sqrt{h}}{\kappa \phi}\phi^{\frac{2}{n}}r+\nonumber \\
&&	(2-n)\frac{1}{\kappa}\int d^{n-1}\by\frac{\sqrt{h}}{\phi}\phi^{\frac{2}{n}}\nabla_i\nabla^i\Omega \ , 
	\nonumber \\
\end{eqnarray}
where we used following formula
\begin{equation}
	\frac{\delta r(\bx)}{\delta h_{ij}(\by)}=-r^{ij}(\bx)\delta(\bx-\by)
	+\nabla^i\nabla^j\delta(\bx-\by)-h^{ij}(\bx)\nabla_k\nabla^k\delta(\bx-\by) \ . 
\end{equation}
In the similar way we obtain 
\begin{eqnarray}
&&	\pb{\bD(\Omega),-
		\int d^{n-1}\by \frac{1}{\kappa}\sqrt{h}\phi^{\frac{2}{n}}
		\frac{(n-1)(n-2)}{n^2\phi^3}h^{ij}\partial_i\phi\partial_j\phi}=\nonumber \\
&&	-\frac{1}{2}\int d^{n-1}\by\Omega 
	\frac{1}{\kappa}\sqrt{h}\phi^{\frac{2}{n}}
	\frac{(n-1)(n-2)}{n^2\phi^3}h^{ij}\partial_i\phi\partial_j\phi
	+\nonumber \\
&&+\int d^{n-1}\by\frac{1}{\kappa}\sqrt{h}\phi^{\frac{2}{n}}
	\frac{(n-1)(n-2)}{n\phi^2}h^{ij}\partial_i\phi\partial_j\Omega \ .  \nonumber \\
\end{eqnarray}
Finally we calculate
\begin{eqnarray}
&&	\pb{\bD(\Omega),\int d^{n-1}\by \frac{2}{\kappa}
		\partial_i[\sqrt{h}h^{ij}\partial_j\phi^{\frac{2}{n}-1}]}=
	\nonumber \\
&&	=-\frac{(n-1)(n-2)}{n}\int d^{n-1}\by\frac{\phi^{\frac{2}{n}}}{\phi^2}
	\partial_i\Omega \sqrt{h}h^{ij}\partial_j\phi-\frac{1}{\kappa}(2-n)\int d^{n-1}\by\phi^{\frac{2}{n}}\sqrt{h}
	\nabla_i\nabla^i\Omega \ .  \nonumber \\	
\end{eqnarray}
Then collecting previous results together we obtain 
\begin{eqnarray}
\pb{\bD(\Omega),\tmH_0(\by)}=0 \ . 
\end{eqnarray}
Finally we  calculate Poisson bracket between $\bD(\Omega)$ and $\mH_i$ and we get
\begin{eqnarray}
	\pb{\bD(\Omega),\mH_i(\by)}
	=-\partial_i\Omega \mD(\by) \ ,  \nonumber \\
\end{eqnarray}
where we used 
\begin{eqnarray}
	\pb{\bD(\Omega),\Gamma_{jm}^k}
	=-\frac{1}{2}\delta^k_m\partial_j\Omega-\frac{1}{2}\partial_m\Omega \delta^k_j
	+\frac{1}{2}h^{kn}\partial_n\Omega h_{jm} \ . \nonumber \\
\end{eqnarray}

With the help of these results we can study time evolution of constraint
$\bD(\Omega)$ and we obtain
\begin{eqnarray}
	\partial_t \bD(\Omega)
=	\pb{\bD(\Omega),H_B}\approx  \frac{1}{\kappa} \int d^{n-1}\bx \pb{\bD(\Omega),\mU}\lambda \phi^{\frac{2}{n}} \approx 0 \ , 
\end{eqnarray}
where we used the fact that there is secondary constraint $\lambda\approx 0$ derived above. 
We see that $\mD\approx 0$ is preserved during time evolution of the system and the requirement of its preservation does not generate additional constraint. 

\subsection{Preservation of Secondary Constraints}
In previous section   we finished analysis of the preservation of the primary constraints during their time evolution. Recall that at this stage $n-1$ secondary constraints 
$\mH_i\approx 0$ was generated. Now we should  check whether these constraints are preserved during time evolution of system.  Again it is convenient to  introduce their smeared form 
\begin{equation}
	\bT_S(\xi^i)=\int d^{n-1}\bx \xi^i\mH_i \ . 
\end{equation}
Using canonical Poisson brackets we obtain
\begin{eqnarray}
&&	\pb{\bT_S(\xi^i),h_{ij}}=
	-\xi^m\partial_m h_{ij}-\partial_i\xi^m h_{mj}-
	h_{im}\partial_j\xi^m \ , \nonumber \\
&&	\pb{\bT_S(\xi^i),\pi^{ij}}=-\xi^m\partial_m\pi^{ij}+
	\partial_m\xi^i\pi^{mj}+\pi^{im}\partial_m\xi^j \ , \nonumber \\
&&	\pb{\bT_S(\xi^i),\phi}=-\xi^m\partial_m\phi \ , \nonumber \\
&&	\pb{\bT_S(\xi^i),p_\phi}=-\partial_i\xi^i p_\phi-\xi^m\partial_m p_\phi
	\nonumber \\
\end{eqnarray}
and consequently  
\begin{eqnarray}
\pb{\bT_S(\xi^i),\tmH_0}=
-\xi^m
\partial_m\tmH_0 \ . 
\nonumber \\
\end{eqnarray}
Then the time evolution of the constraint $\bT_S(\xi^i)$ is equal to
\begin{eqnarray}\label{premHi}
&&	\partial_t \bT_S(\xi^i)=
	\pb{\bT_S(\xi^i),H_B}=
	\nonumber \\
&&	=\int d^{n-1}\bx(-\omega \xi^m\partial_m\mH_0+\frac{1}{\kappa}\lambda
\phi^{\frac{2}{n}} \pb{\bT_S(\xi),\mU}+\pb{\bT_S(\xi^i),N^i\mH_j})\approx\nonumber \\
&&	\approx -\int d^{n-1}\bx  \omega \xi^m\partial_m\mH_0\approx 0 \ ,  \nonumber \\
\end{eqnarray}
where we also used the fact that Poisson bracket between smeared forms of spatial diffeomorphism constraints is equal to
\begin{eqnarray}
\pb{\bT_S(\xi^i),\bT_S(\zeta^j)}=\bT_S(\xi^m
\partial_m\zeta^n-\zeta^m\partial_m\xi^n) \ 
\nonumber \\
\end{eqnarray}
and hence it vanishes on the constraint surface $\mH_m\approx 0$.

The equation (\ref{premHi}) shows that spatial diffeomorphism constraints are preserved during the time evolution of the system on condition when we impose 
 tertiary constraint
\begin{equation}\label{partmH0}
\partial_m\tmH_0 \approx 0 \ . 
\end{equation}
We can deal with this equation in the following way. Let us split $\tmH_0$ into its zero mode part and the remaining one
\begin{equation}\label{tmH0split}
	\tmH_0=\frac{1}{\int d^{n-1}\bx \omega}\bH(t)+\hat{\mH}_0 \ ,  \quad 
	\bH=\int d^{n-1}\bx\omega \tmH_0 \ , 
\end{equation}
where $\hat{\mH}_0$ obeys condition 
\begin{equation}
	\int d^{n-1}\bx \omega \hat{\mH}_0=0  
\end{equation}
as follows from definition (\ref{tmH0split}). Then (\ref{partmH0}) 
does not restrict $\bH$ while there   are $(n-1)\times 
\infty-1$ constraints
\begin{equation}
	\hat{\mH}_0\approx 0 \ . 
\end{equation}
Before we will discuss 
separation of $\tmH_0$ in more details  we should determine Poisson brackets between $\tmH_0(\bx)$ and $\tmH_0(\by)$. 
It is again convenient to introduce smeared form of 
$\tmH_0$
\begin{equation}
	\bT_T(X)=\int d^{n-1}\bx \omega X \tmH_0 \ .
\end{equation}	
Then after some calculations that are similar to the calculations when
we checked stability of constraint $\mD\approx 0$ we obtain 
\begin{eqnarray}\label{pbHam}
&&	\pb{\bT_T(X),\bT_T(Y)}=
	\bT_S((X\partial_iY-Y\partial_iX)h^{ij}\frac{\phi^2\omega^2}{h})+\nonumber \\
&&	
	+\bD(\frac{2}{n-1}(Y\partial_l\partial_k X-X\partial_l\partial_kY)h^{kl}\frac{\phi^2\omega^2}{h}) 
	\nonumber \\
\end{eqnarray}
which shows that Poisson brackets between smeared form of $\tmH_0$ vanish on the constraint surface $\mH_i\approx 0 \ , \mD\approx 0$. 
Now we can return to the definition of constraint $\hat{\mH}_0$. We introduce its smeared form 
\begin{equation}
	\hat{\bT}_T(X)=\int d^{n-1}\bx\omega \hat{X} \hat{\mH}_0 \ , 
\end{equation}
where $\hat{X}$ obeys the condition $\int d^{n-1}\bx \hat{X}\omega=0$. 
Then we have
\begin{eqnarray}
&&\pb{\hat{\bT}_T(\hat{X}),\bH}=\pb{\int d^{n-1}\bx \omega X
	\hat{\mH}_0,\bH}=\nonumber \\
&&=\pb{\int d^{n-1}\bx\omega \hat{X}(\tmH_0-\frac{1}{V}\bH),\bH}=\nonumber \\
&&=-\bT_S(\partial_iX h^{ij}\frac{\phi^2\omega^2}{h})+
\bD(\frac{2}{n-1}(\partial_l\partial_k Xh^{kl}\frac{\phi^2\omega^2}{h}))\approx 0 .
\nonumber \\
\end{eqnarray}
Finally we will calculate Poisson bracket of smeared form of the constraints $
\hat{\mH}_0$ using the Poisson brackets (\ref{pbHam}) 
\begin{eqnarray}
&&\pb{\hat{\bT}_T(\hat{X}),\hat{\bT}_T(\hat{Y})}=
\bT_S((\hat{X}\partial_i\hat{Y}-\hat{Y}\partial_i\hat{X})h^{ij}
\frac{\phi^2\omega^2}{h})+\nonumber \\
&&+\bD(\frac{2}{n-1}(\hat{Y}\partial_{l}\partial_k\hat{X}-
\hat{X}\partial_l\partial_k\hat{Y})
h^{kl}\frac{\phi^2\omega^2}{h})
\nonumber \\
\end{eqnarray}
that shows that $\hat{\mH}_0$ are first class constraints. However it is important to stress an existence of global mode $\bH$. In fact, the time evolution of $\bH$ is equal to
\begin{equation}\label{HB}
\frac{d\bH}{dt}=\pb{\bH,H_B}\approx \int d^{n-1}\bx \omega\frac{1}{\kappa}\lambda\phi^{\frac{2}{n}} \pb{\bH,\mU}=0
\end{equation}
using the fact that $H_B=\bH+\bT_S(N^i)+\int d^{n-1}
\bx (\frac{1}{\kappa}\lambda \omega \phi^{\frac{2}{n}}\mU+v_N\pi_N+v^i\pi_i+v^\lambda
p_\lambda)+\bD(\Sigma)$ 
and the fact that $\bD(\Omega)$ is the first class constraint. 

The result (\ref{HB}) shows 
that $\bH=\mathrm{const}$ and corresponds to the cosmological constant of theory which is the same situation as in the case of unimodular gravity. In fact, we show in the next section how unimodular gravity arises from WTG by gauge fixing
 Weyl symmetry. 

\subsection{Gauge Fixing of Weyl Symmetry}
As we argued in previous section Weyl transverse gravity possesses gauge symmetry which is Weyl rescaling of metric. Reflection of this fact is the presence of the first class constraint $\mD\approx 0$. We can fix this symmetry by introducing gauge fixing function. In principle there are many such functions where the only requirement that this function has to obey  is that it has to have non-zero Poisson bracket with $\mD$. We choose the simplest one when the gauge fixing function $\mF$ has the form 
\begin{equation}
\mF\equiv \phi-1\approx 0 \ . 
\end{equation}
 It is easy so see that $\pb{\mF,\mD}=\frac{n}{2}\phi \neq 0 $ and hence $\mF$ is legitimate gauge fixing function. Then $\mF$ and $\mD$ are second class constraints
that are strongly zero where $\mF$ means that $\phi=1$ while from $\mD=0$ we obtain $p_\phi$ to be equal to
\begin{equation}
p_\phi=-\frac{2}{n}\pi \ . 
\end{equation}
Inserting this relation to the definition of $\tmH_0$ we obtain gauge fixed form of $\tmH_0$
\begin{eqnarray}\label{fixed}
\tmH_0(fixed)
=\frac{\kappa}{h}(\pi^{ij}h_{ik}h_{jl}-\frac{1}{n-2}\pi^2) 
-\frac{1}{\kappa }r 
\nonumber \\ 
\end{eqnarray}
using  the fact that $\Pi^{ij}(p_\phi=-\frac{2}{n}\pi)$ is equal to
\begin{equation}
\Pi^{ij}(p_\phi=-\frac{2}{n}\pi)=\pi^{ij}-\frac{1}{n-1}\pi h^{ij} \ . 
\end{equation}
 It is important to stress that we fixed the  gauge  
\emph{after} complete canonical analysis of WTG was performed
with explicit identification of the zero mode part of Hamiltonian as
cosmological constant and we also identified $(n-1)\times \infty-1$ $\hat{\mH}_0\approx 0$ as first class   constraints.  
 In other words we have explicitly shown that gauge fixing 
in Weyl transverse gravity leads to unimodular theory of gravity which 
is nice consistency check.

{\bf Acknowledgement:}

This work  is supported by the grant “Dualitites and higher order derivatives” (GA23-06498S) from the Czech Science Foundation (GACR).

\end{document}